\documentclass[letterpaper]{article} 
\usepackage{aaai23}  
\usepackage{times}  
\usepackage{helvet}  
\usepackage{courier}  
\usepackage[hyphens]{url}  
\usepackage{graphicx} 
\usepackage{natbib}  
\usepackage{caption} 
\usepackage{algorithm}
\usepackage{algorithmic}

\usepackage{tikz}
\usepackage{comment}
\usepackage{amsmath,amssymb} 
\usepackage{color}
\usepackage{booktabs}
\usepackage{adjustbox}
\usepackage{makecell}
\usepackage{xcolor}
\usepackage{adjustbox}
\usepackage{multirow}
\usepackage{epsfig,epstopdf}
\usepackage{bbding}
\usepackage{balance}
\usepackage[capitalize]{cleveref}
\crefname{section}{Sec.}{Secs.}
\Crefname{section}{Section}{Sections}
\Crefname{table}{Table}{Tables}
\crefname{table}{Tab.}{Tabs.}

\usepackage{newfloat}
\usepackage{listings}
\DeclareCaptionStyle{ruled}{labelfont=normalfont,labelsep=colon,strut=off} 
\floatstyle{ruled}
\newfloat{listing}{tb}{lst}{}
\floatname{listing}{Listing}
\title{Hybrid Pixel-Unshuffled Network for Lightweight Image Super-Resolution}
\author {
   Bin Sun\textsuperscript{\rm 1,\rm 3}, 
   Yulun Zhang\textsuperscript{\rm 2}, 
   Songyao Jiang\textsuperscript{\rm 1}, 
   Yun Fu\textsuperscript{\rm 1,\rm 3}
}
\affiliations {
    \textsuperscript{\rm 1} Northeastern University, Boston, MA, USA, 
    \textsuperscript{\rm 2} ETH Z\"{u}rich, Z\"{u}rich, Switzerland \\
    \textsuperscript{\rm 3}AInnovation Labs Inc., Boston, MA, USA \\
    sun.bi@northeastern.edu, \{yulun100, jiangsongyao\}@gmail.com, yunfu@ece.neu.edu
}

\begin{document}

\maketitle

\begin{abstract}
Convolutional neural network (CNN) has achieved great success on image super-resolution (SR). However, most deep CNN-based SR models take massive computations to obtain high performance. Downsampling features for multi-resolution fusion is an efficient and effective way to improve the performance of visual recognition. Still, it is counter-intuitive in the SR task, which needs to project a low-resolution input to high-resolution. In this paper, we propose a novel Hybrid Pixel-Unshuffled Network (HPUN) by introducing an efficient and effective downsampling module into the SR task. The network contains pixel-unshuffled downsampling and Self-Residual Depthwise Separable Convolutions.
Specifically, we utilize pixel-unshuffle operation to downsample the input features and use grouped convolution to reduce the channels. Besides, we enhance the depthwise convolution's performance by adding the input feature to its output.
The comparison findings demonstrate that, with fewer parameters and computational costs, our HPUN achieves and surpasses the state-of-the-art performance on SISR. All results are provided in the github link\footnote{https://github.com/Sun1992/HPUN}.

\end{abstract}

\setlength{\abovedisplayskip}{2pt}
\setlength{\belowdisplayskip}{2pt}
\section{Introduction}

Single Image Super-Resolution (SISR) is a fundamental vision task which involves accurately reconstructing a high-resolution (HR) image from a low-resolution (LR) image.
SISR has been utilized on various high-level tasks such as face synthesis~\cite{yin2020joint,yin2021superfront}, medical imaging~\cite{shi2013cardiac}, surveillance imaging~\cite{zou2012very}, and image generation~\cite{karras2018progressive}. Dong \textit{et al}.~\cite{dong2014learning} first introduced CNN into SISR and achieved impressive performance in 2014. Afterward, more deep CNN methods are proposed for the super-resolution tasks~\cite{schulter2015fast,huang2015single,kim2016accurate,kim2016deeply,lim2017enhanced,tong2017image,tai2017memnet,zhang2018learning,rcan2018}. Among these, one of the most fundamental architectures is EDSR~\cite{lim2017enhanced}. However, these networks need expensive computation resources, which is the main bottleneck for their deployment on mobile devices for real-time purposes.
\begin{figure}[!t]
\setlength{\abovecaptionskip}{-0.1 cm}
\setlength{\belowcaptionskip}{-0.3cm}
\centering
\includegraphics[width=1.0\linewidth]{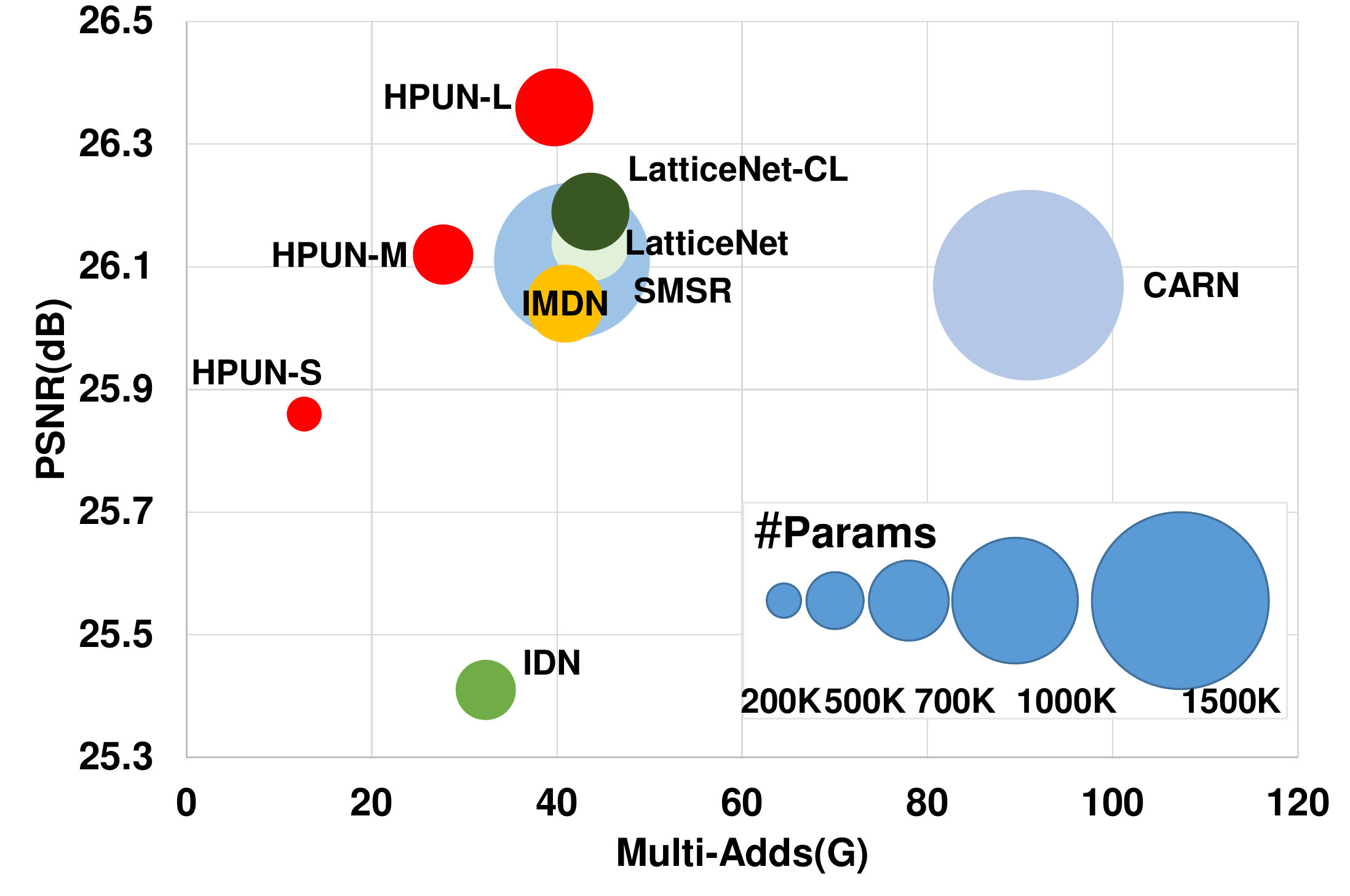}
\caption{
Illustration of the overall comparison on Urban100 with $\times4$ scale. Our proposed HPUN-L achieves the best trade-off among the PSNR, parameters, and Multi-Adds. 
}\label{fig:overall_comparison}
\vspace{-0.35cm}
\end{figure}

\begin{figure*}[t]
\centering
\includegraphics[width=0.9\linewidth]{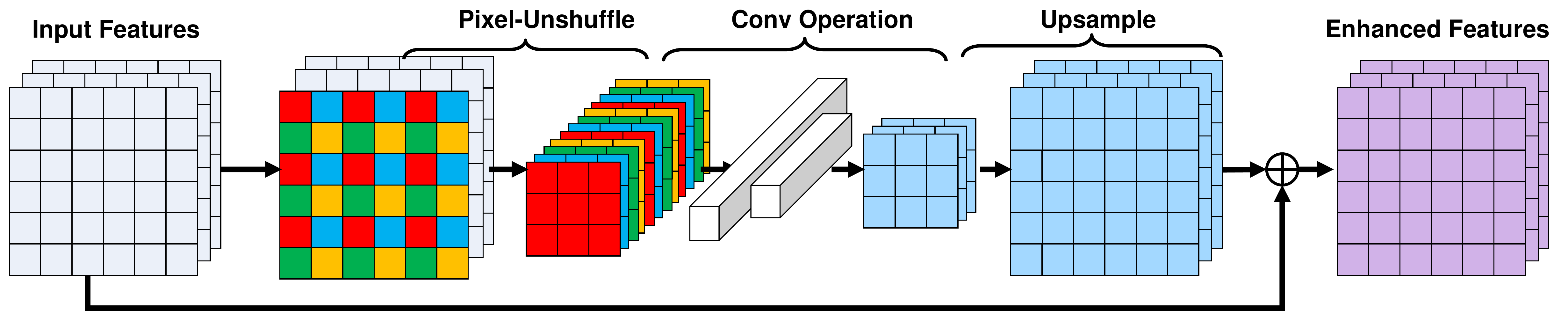}
\caption{The scheme of our proposed pixel-unshuffled downsampler. Note that the notation of Conv Operation in the figure is a general operation. We will explore the best operations in the following sections.}
\label{fig:scheme} 
\vspace{-5.5mm}  
\end{figure*}
In recent years, a lot of manually designed lightweight structures have been proposed~\cite{sifre2014depthconv,mobilenets,chollet2017xception,iandola2016squeezenet,Mobilenetv2,Zhang_2018_shufflenet,shufflenetv2,mobilenetv3,mixnet2019,han2020ghostnet}. Among these structures, the most fundamental one is the depthwise convolution layer~\cite{sifre2014depthconv}, which processes the spatial information with a single convolution on each input feature. A $1\times1$ convolution layer named pointwise layer is usually deployed around the depthwise convolution layer for the communication among channels~\cite{mobilenets,Mobilenetv2,mobilenetv3,Zhang_2018_shufflenet,shufflenetv2}. However, such structures are not popular in the SISR due to their significant performance loss. CARN~\cite{CARN2018} tried to use a similar structure as MobileNet~\cite{mobilenets} on SISR in 2018. They utilized the group convolution to reduce the parameters, but they had to introduce a complicated recurrent scheme to improve the performance. As shown in Figure~\ref{fig:overall_comparison}, the computation costs and parameters of CARN are not satisfied. Therefore, it is still a main challenge to effectively implement depthwise convolution based lightweight structures to the image super-resolution task.  

Besides using lightweight operations, the computation costs can be alleviated by reducing the size of feature maps~\cite{tan2019efficientnet,mobilenets,chollet2017xception,iandola2016squeezenet,Mobilenetv2,Zhang_2018_shufflenet,shufflenetv2,mobilenetv3,mixnet2019,han2020ghostnet}. 
Meanwhile, size-reduced features can also improve high-level representations by merging with higher-resolution features in many tasks~\cite{HRnet2019,HRNetTPAMI}. However, it is counter-intuitive to apply downsampling modules in SISR since SISR is an upsampling task that restores information of a low-resolution image. In contrast, the downsampling operation usually causes significant information loss. 
To direct the reconstruction, Haris \textit{et al}.~\cite{haris2018deep} suggested an iterative error-correcting feedback method that computes both up- and down-projection errors.
Furthermore, Li \textit{et al}.~\cite{SRFBN-S2019} also proposed a framework that introduced the downsampling module into SISR to generate high-level representations. Their success shows the possibility of getting pleasing high-resolution images through downsampling operations. However, they still had to utilize a recurrent scheme to resist the performance drop, which heavily increased the parameters and computation costs. 


This paper explores an effective way to design a lightweight network with depthwise convolutions and downsampling operations. Specifically, we develop a simple yet effective module named Self-Residual Depthwise Separable Convolution to overcome the drawback in Depthwise Separable Convolution (DSC)~\cite{mobilenets} without any additional parameters. Inspired by the previous explorations on downsampling features~\cite{HRnet2019,haris2018deep,SRFBN-S2019,gu2019self}, we propose a pixel-unshuffled downsampler constructed with the pixel-unshuffle operation, max-pooling, and group convolution to further enhance the performance of DSC with similar computation costs as depthwise convolution. Specifically, the pixel-unshuffle operation is the reverse operation of pixel-shuffle~\cite{shi2016real}, which can help to avoid information loss shown by Gu \textit{et al}.~\cite{gu2019self} The scheme is shown in Figure~\ref{fig:scheme}. Moreover, we propose a practical, lightweight module named Pixel-Unshuffled Block (PUB) constructed with the pixel-unshuffled downsampling and the Self-Residual DSC. Last, we replace one Self-Residual DSC in the PUB with a standard convolution layer and construct a Hybrid Pixel-Unshuffled Network (HPUN) to achieve the state-of-the-art (SOTA) performance and slightly increase the number of the HPUN to beyond the SOTA performance. The overall comparison is shown in Figure~\ref{fig:overall_comparison}.
The main contributions are summarized as:
\begin{itemize}
\item We propose the Self-Residual DSC to overcome the defects of the depthwise convolution in the SISR task with a simple and effective operation, which barely needs computation and additional parameters.
\item We propose a novel downsampling module with the pixel-unshuffle operation, which is our key module to enhance the performance. 
\item We propose a lightweight module named PUB with our Self-Reisdual DSC and the pixel-unshuffled downsampler, which can provide reliable performance with a few parameters and computation costs.
\item We propose the Hybrid Pixel-Unshuffled Block (HPUB) by integrating the standard convolution into the PUB, and construct the effective and efficient HPUN to achieve a new SOTA performance with a few parameters and Multi-Adds comparing the baselines.
\item We discover the relationship between PSNR and the Normalized Mean Error (NME) among the shallow features and deep features based on our ablation study, which may be valuable in designing the network for SISR. We will discuss the details in the experiment section. 

\end{itemize}

\section{Related Work}
\noindent\textbf{Deep Super Resolution.} SRCNN developed the initial end-to-end system that converts the interpolated LR images to their HR counter-parts~\cite{dong2014learning}. 
The SRCNN was further improved by its successors with advanced network architectures~\cite{kim2016accurate,zhang2017learning}. As studied in~\cite{dong2016accelerating}, computational costs are quadratically increased by this upsampling operation in data preprocessing. To solve the problem, an efficient sub-pixel convolution layer that upsampled the last LR feature maps to HR was introduced in ESPCN~\cite{shi2016real}. It was also adopted in residual-learning networks SRResNet~\cite{ledig2017photo} and EDSR~\cite{lim2017enhanced}. The performance of the SISR was then further improved by stacking more blocks with dense residuals~\cite{zhang2017residual,zhang2018densely,zhang2018residual,zhang2019rnan}.

\noindent\textbf{Lightweight Super Resolution.} LapSRN~\cite{lai2017deep} reduced the computation complexity by removing the bicubic interpolation before prediction. Inspired by LapSRN, a lot of works started to reconstruct the HR image from the origin LR input. Recursive learning was first introduced by DRCN~\cite{kim2016deeply}. Then it was widely used to reduce the parameters with weight sharing strategy~\cite{tai2017image,tai2017memnet,haris2018deep,CARN2018,SRFBN-S2019}. Besides the recurrent scheme, IDN~\cite{IDN2018} and CARN~\cite{CARN2018} introduced the group convolution for the lightweight purpose. As the success of the residual operation in SISR, many works~\cite{IDN2018,hui2019lightweight,latticenet2020} adopted the residual into their lightweight design to keep the performance. A recent work named SMSR~\cite{SMSR2021} reduced the parameters and computation costs with pruning. Different with SMSR, we manually design the lightweight network which can be further improved by pruning.

\begin{figure}[!t]
\setlength{\abovecaptionskip}{0.3cm}
\setlength{\belowcaptionskip}{-0.3cm}
\centering
\includegraphics[width=0.7\linewidth]{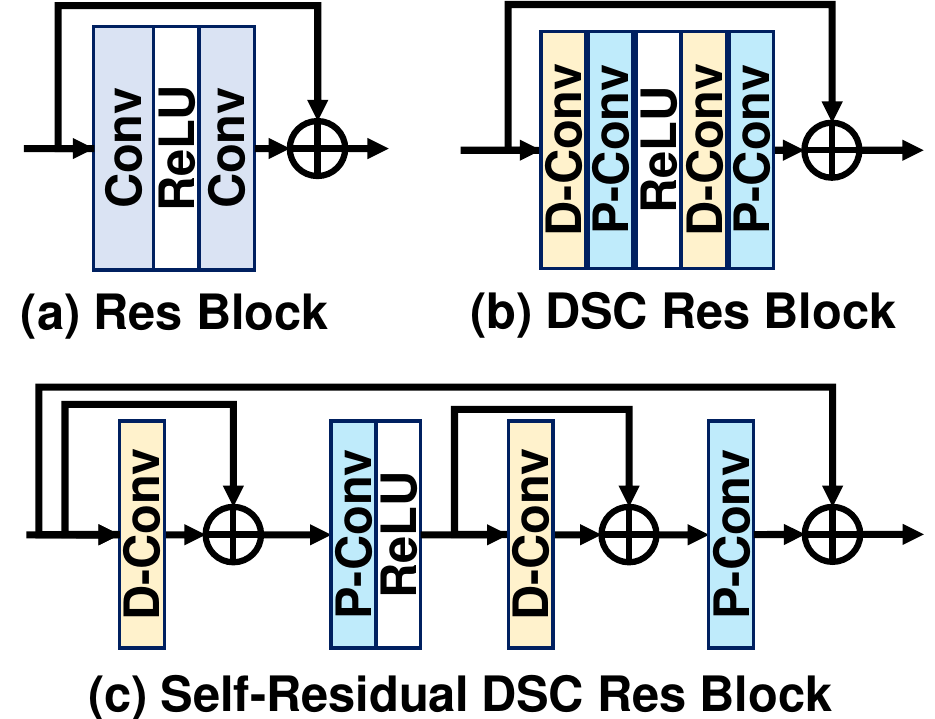}
\caption{
Illustration of modules. (a) standard residual block. (b) the residual block constructed by the DSC. (c) the residual block constructed by our proposed Self-Residual DSC. Abbrevs.: D-Conv=Depthwise Convolution, P-Conv=Pointwise Convolution.
}
\label{fig:DSCmodule}
\vspace{-3mm}
\end{figure}

\noindent\textbf{Deep Lightweight Structure.} In recent years, as the deep-learning models become deeper and larger, many researchers have been working on the lightweight networks.
An activation function named ReLU was proposed for speed-up purpose~\cite{relu}. 
There was a flattened CNN architecture that sped up the forward feeding presented in~\cite{jin2014flattened}. 
The concept of depthwise separable convolution was initially proposed by Sifre \textit{et al}.~\cite{sifre2014depthconv} and was widely adopted in Inception~\cite{ioffe2015batch}, Xception~\cite{chollet2017xception},  ShuffleNets~\cite{Zhang_2018_shufflenet,shufflenetv2}, MobileNets~\cite{mobilenets,Mobilenetv2}, CondenseNet~\cite{condensenet}, and LRPRNet~\cite{lrpr}.
Besides, researchers suggested using Neural Architecture Search (NAS) in addition to manually created lightweight architectures to locate the ideal lightweight network~\cite{darts2019,NasNet2018,proxylessnas2019,fbnet2019,mobilenetv3,mixnet2019}. All these networks are constructed based on the depthwise convolution as well. Thus, it is necessary to explore an effective way to implement the depthwise convolution on SISR. In this work, we provide a module that can greatly improve the depthwise convolution's performance for super resolution tasks.

\section{Proposed Method}
We propose a lightweight structure called Hybrid Pixel-Unshuffled Block to replace the traditional Residual Convolution Block, which is shown in Figure~\ref{fig:DSCmodule}(a). Our proposed method has three parts: a standard convolution layer, the proposed pixel-unshuffled downsampling, and the proposed PUB. Specifically, the PUB is an integration of the pixel-unshuffled downsampling and the Self-Residual DSC. Therefore, we organize this section as follows: first, we will introduce the details of the Self-Residual DSC; the pixel-unshuffled downsampling will be introduced in the second sub-section; at last, we will present the details of the HPUN.
\subsection{Self-Residual DSC}
\label{sec:DSC}
\noindent\textbf{DSC.} Depthwise separable convolution (DSC) is composed by a depthwise layer and a pointwise layer as shown in Fig.~\ref{fig:DSCmodule}(b). The depthwise layer uses single kernel for each input feature map. 
DSC is a popular lightweight module to reduce the redundant operations in the standard convolution. The conversion from the standard convolution to the DSC can be described as:
\begin{equation}
\mathbf{F}^{\text{out}} = C(\mathbf{F}^{\text{in}}) \approx P(D(\mathbf{F}^{\text{in}})),
\label{equ.0}
\end{equation}
where $\mathbf{F}^{\text{out}}$ means the output features, $C$ represents the standard convolution, $\mathbf{F}^{\text{in}}$ means the input features, $D$ stands for the depthwise convolution, while $P$ stands for the pointwise convolution. Depthwise convolution is the major part to process the spatial information of the input features, which needs far fewer parameters and computation costs than standard convolution with the same kernel settings.

\vspace{1mm}
\noindent\textbf{Self-Residual DSC.} It may have a significant side effect on the performance of SISR since SISR needs to enrich the information. The side effect is shown in the experiment section. To overcome the defects brought by the depthwise layer and keep its ability to process the spatial information, We find a balanced trade-off design by simply adding the input before the depthwise layer to the output of the depthwise layer as shown in Figure~\ref{fig:DSCmodule}(c). The whole structure is described as:
\begin{equation}
\mathbf{F}^{\text{out}} = P(D(\mathbf{F}^{\text{in}})+\mathbf{F}^{\text{in}}).
\label{equ.5}
\end{equation}
Comparing Equation~\eqref{equ.0} and Equation~\eqref{equ.5}, we can easily figure out that the outputs of the Self-Residual DSC have more similarity to the inputs than the outputs of the DSC. We will analyze the importance of the similarity in the experiment section. The self-residual does not introduce any additional parameters, and its computation costs can be ignored.

\subsection{Pixel-Unshuffled Downsampling}

\begin{figure}[!t]
\setlength{\abovecaptionskip}{-0.0cm}
\setlength{\belowcaptionskip}{-0.2cm}
\centering
\includegraphics[width=0.7\linewidth]{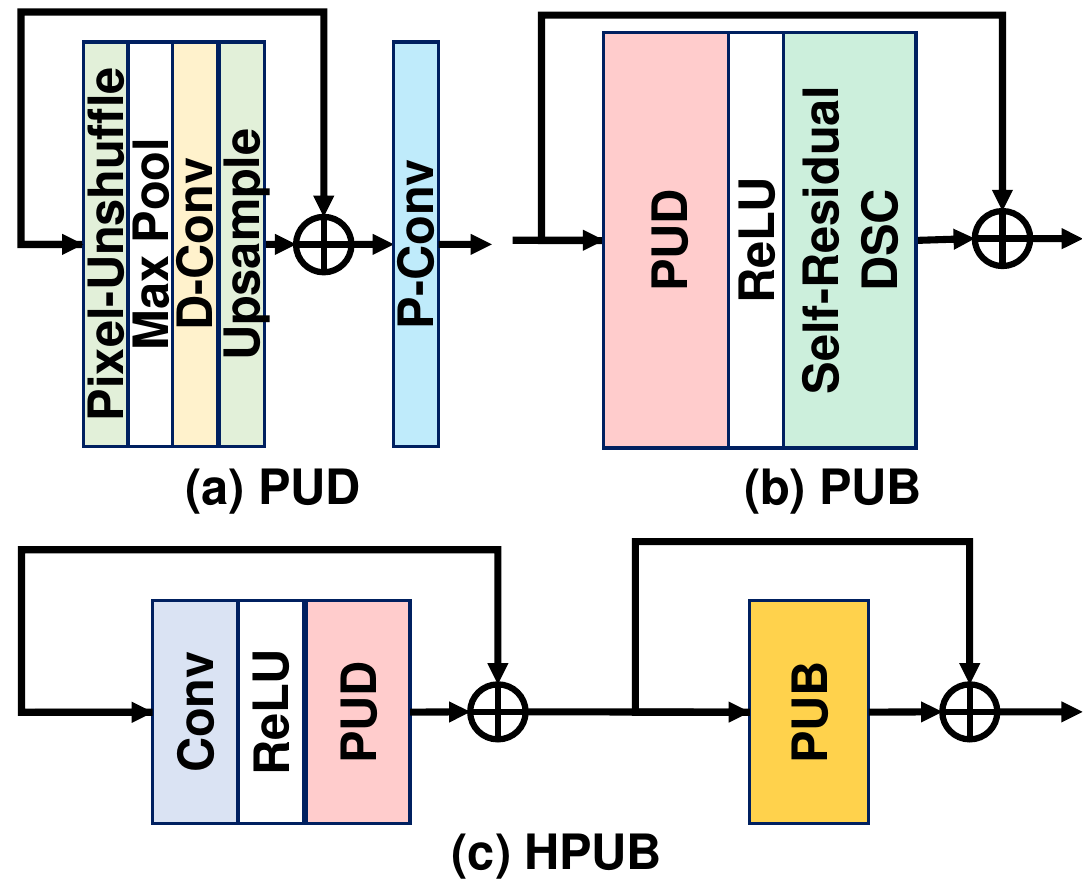}
\caption{ (a) Pixel-Unshuffled Downsampling (PUD). (b) Pixel-Unshuffled Block (PUB). (c) Hybrid Pixel-Unshuffled Block (HPUB). Note that the group number of the depthwise convolution in the PUD equals to the number of its inputs. 
}
\label{fig:pudmodule}
\vspace{-3mm}
\end{figure}

Now we present details about the pixel-unshuffled downsampling, which is shown in Figure~\ref{fig:pudmodule}(a). As discussed in previous sections, low-frequency features can enhance the high-level representations~\cite{HRnet2019,HRNetTPAMI}. 
The work~\cite{HRNetTPAMI} investigates that repeating multi-resolution fusions can improve high-resolution representations for image segmentation tasks by leveraging low-resolution representations.
However, previous SR works~\cite{haris2018deep,SRFBN-S2019} took a lot of effort to use the low-frequency features in SISR with a heavy recurrent scheme. In this work, we explore a more efficient way to utilize the low-frequency features with single forward inference for the SISR task. The proposed scheme is shown in Figure~\ref{fig:scheme}. In this subsection, we will first introduce the pixel-unshuffle operation. Then we will focus on exploring the most effective operations after the pixel-unshuffle. 

\noindent\textbf{Pixel-unshuffle.} Pixel-unshuffle is a reverse operation of pixel-shuffle~\cite{shi2016real}. As shown in Figure~\ref{fig:scheme}, it divides a feature into several sub-features, whose number in this work is four. We use four different color to represent sub-features in the figure. As shown in the figure, the sub-features contain the complete information of the original features but with lower resolution. Therefore, we use it to avoid information loss while reducing the size of the features, which can be processed by some complicated operations with lower consumption.

\noindent\textbf{Pixel-Unshuffled Downsampling.} After the pixel-unshuffle operation, we need a efficient and effective structure to process the low-resolution features. As shown in Figure~\ref{fig:pudmodule}(a), we first deploy a max-pooling after the pixel-unshuffle. The reason is we want a powerful non-linear operation before the convolution operation to extract better local features. Therefore, we choose the max-pooling instead of the average-pooling layer. We describe the process as:
\begin{equation}
F^{\text{out}}_{i,j} = M(F^{\text{in}}_{i,j}), \quad i\in\{1,2,3,4\}, j\in\{1 \dots n\},
\label{equ.1}
\end{equation}
where $F^{\text{in}}_{i,j}$ means the $i$th sub-feature of $j$th input channel, $F^{\text{out}}_{i,j}$ means the $i$th sub-feature of $j$th output channel, and $M$ means max-pooling with stride $1$. 

After the non-linear operation, we use a group convolution to reduce the channel of the input, which is actually a downsampling operation. The process can be described as:
\begin{equation}
F^{\text{out}}_{j} = G(F^{\text{in}}_{1,j},F^{\text{in}}_{2,j},F^{\text{in}}_{3,j},F^{\text{in}}_{4,j}), \quad j \in \{1 \dots n\},
\label{equ.2}
\end{equation}
where $F^{\text{out}}_{j}$ means the $j$th output channel, and $G$ means the grouped convolution whose group number equals to $4$. 

To enhance the feature, we use an upsampler to project the low-frequency features to high dimension, and add them to the original input features. After that, a pointwise convolution is utilized for the communication among the channels. The process can be described as:
\begin{equation}
\mathbf{F}^{\text{out}} = P(U(\mathbf{F}^{\text{in}})+\mathbf{L}),
\label{equ.3}
\end{equation}
where $U$ stands for the upsampling function, $\mathbf{F}^{\text{in}}$ means the input channels to the upsampler, and $\mathbf{L}$ means the original input features. We use bi-linear upsampler. 
\subsection{Hybrid Pixel-Unshuffled Network}
\noindent\textbf{Pixel-Unshuffled Block.} After the exploration of the Self-Residual DSC and the pixel-unshuffled downsampling, we introduce the lightweight Pixel-Unshuffled Block (PUB). The PUB is composed of the Self-Residual DSC and the pixel-unshuffled downsampling. The detail of PUB is shown in Figure~\ref{fig:pudmodule}(b). The block can be represented as:
\begin{equation}
\mathbf{F}^{\text{out}} =P(D(\sigma(\text{PUD}(\mathbf{F}^{\text{in}})))+\sigma(\text{PUD}(\mathbf{F}^{\text{in}})))+\mathbf{F}^{\text{in}},
\label{equ.4}
\end{equation}
where the $\text{PUD}$ denotes the whole procedure of the pixel-unshuffled downsampling, and $\sigma$ represents the ReLU~\cite{relu}.

\vspace{1mm}
\noindent\textbf{Hybrid Pixel-Unshuffled Block.}
To further improve the performance on reconstruction, we integrate the standard convolution into our proposed PUB, which is named Hybrid Pixel-Unshuffled Block (HPUB). The details of the HPUB are shown in Figure~\ref{fig:pudmodule}(c). The standard convolution layer's kernel size is set to $3$ to strike a balance between efficiency and performance. The kernel settings for the rest modules are the same as the PUB.

\begin{figure}[t]
\centering
\includegraphics[width=1.0\linewidth]{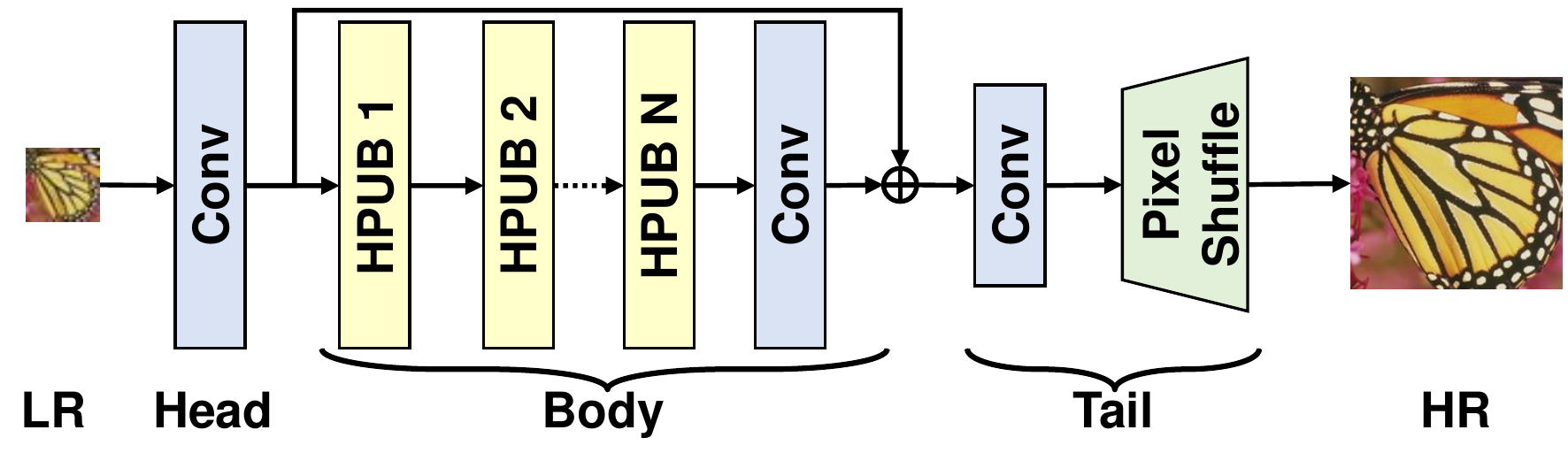}
\caption{Our proposed HPUN. The architecture is based on EDSR, and the tail is from IMDN. The total number of HPUBs in HPUN is no less than $8$.}
\label{fig:HPUN} 
\vspace{-5mm} 
\end{figure}
\vspace{1mm}
\noindent\textbf{Hybrid Pixel-Unshuffled Network.}
We use HPUB to construct our Hybrid Pixel-Unshuffled Network (HPUN). The network is similar to EDSR. Since one HPUB has two residual blocks, we construct the body parts with $8$ HPUB to align the settings in EDSR. To further reduce the parameters, we use the tail of IMDN. The details of the architecture are shown in Figure~\ref{fig:HPUN}. We control the total number of HPUB for different sizes of the model. HPUN-M has 8 HPUN blocks in total, and HPUN-L has 12 HPUN blocks. The total number of HPUN-S is 8, but we replace one HPUB with two PUBs. The upsampler for the final high-resolution output is the pixel-shuffle module~\cite{shi2016real}.

\vspace{-1mm}
\section{Experimental Results}
\label{sec:results}
\label{subsec:settings}
\noindent\textbf{Datasets and Metrics.} 
As training data, we use the DIV2K dataset~\cite{timofte2017ntire} following the popular works~\cite{han2015learning,timofte2017ntire,lim2017enhanced,zhang2018learning}. We used the following testing datasets: Set5~\cite{bevilacqua2012low}, Set14~\cite{zeyde2012single}, B100~\cite{martin2001database}, Urban100~\cite{huang2015single}, and Manga109~\cite{matsui2017sketch}. PSNR and SSIM~\cite{wang2004image} on the Y channel (\textit{i.e.} luminance) of transformed YCbCr space are the evaluation metrics. The degradation is bicubic (denote as \textbf{BI} for short) downsampling implemented with the Matlab function \textit{imresize}~\cite{zhang2018learning,zhang2019rnan}.
We use \textbf{BI} mode to simulate LR images. The scaling factors are $\times2$, $\times3$, and $\times4$. We also compare the parameters and Multi-Adds to evaluate the spatial and time complexity. The results of $\times2$ and $\times3$ will be shown in the supplementary material.


\vspace{1mm}
\noindent\textbf{Training Setting.}
Following the popular settings~\cite{lim2017enhanced}, we extract 16 LR RGB patches at random as inputs in each training batch.
The size of each patch is $48\times48$. The patches are randomly augmented by flipping horizontally or vertically and rotating 90$^{\circ}$. There are 14,200 iterations in one epoch. We implement our HPUN with the PyTorch~\cite{Pytorch} and update it with Adam optimizer~\cite{kingma2014adam}. The learning rate is initialized to $2\times10^{-4}$ for all layers and follows the cosine scheduler with 250 epochs in each cycle. We finetune the model with longer epochs and larger batchsize for final comparisons. Some experiments use the step scheduler and will be emphasized in the caption for fair comparison.

\subsection{Ablation Study}

\label{secablationstudy}
We first demonstrate the effectiveness of the Self-Residual DSC. Then we will show the enhancement of the pixel-unshuffled downsampling. A set of experiments are implemented to figure out the best operation in the pixel-unshuffled downsampling. At last, we will visualize the features and discuss our intuition. More results of the ablation study will be shown in the supplementary material.

\begin{table}[t]
\begin{center}
\caption{Experiment results of different components generated with step schedule. The kernel size in PUD is set to $3$ for fair comparison. PUB is the combination of PUD and Self-Residual DSC. }
\scalebox{0.8}{
\begin{tabular}{|l | c| c| c|c|}
\hline
Methods &PSNR(Set5)& Params & Multi-Adds \\
 \hline
P-Conv & 37.36 & 181K & 40.7G\\
\hline
DSC  & 37.65 & 207K & 44.9G\\
\hline

Self-Residual DSC & 37.73 & 207K & 44.9G\\
\hline
PUD & 35.68 & 267K & 44.9G\\
\hline
PUB & 37.76 & 226K & 44.9G\\
\hline
\end{tabular}}
\vspace{-3mm}

\label{table:ImprovedDSC}
\end{center}
\end{table}



\begin{figure}[t]

\centering
\includegraphics[width=0.8\linewidth]{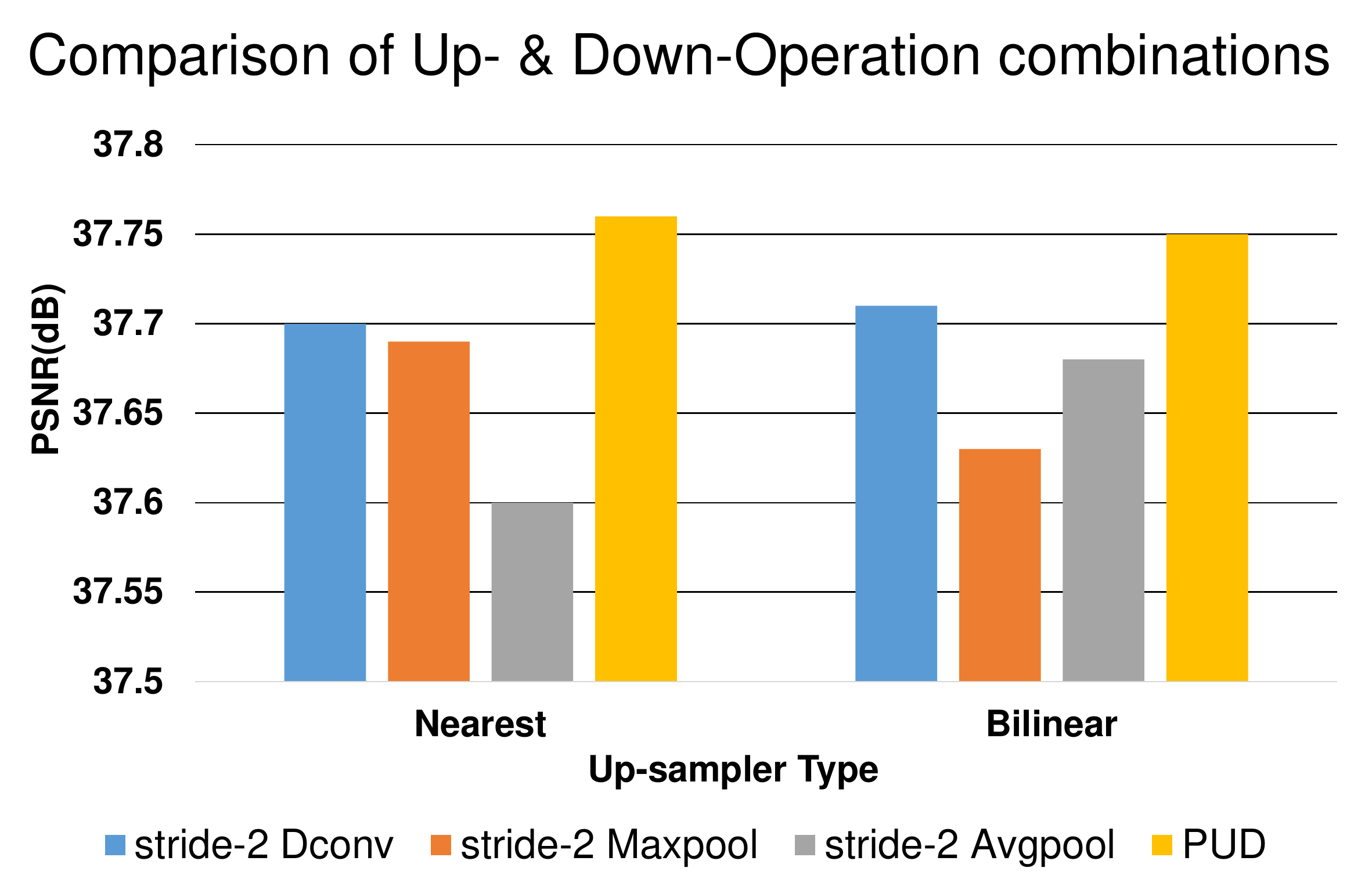}
\caption{Results of different combinations of up- and down-scale operations conducted with step scheduler. }
\label{fig:Kernel_size} 
\vspace{-6mm}  
\end{figure}

\noindent\textbf{Effectiveness of the Self-Residual DSC.}  
From the Table~\ref{table:ImprovedDSC}, we can find out that the depthwise convolution is necessary for lightweight architecture since the P-Conv cannot process the spatial information in the features. The DSC achieve around $0.3$ dB higher PSNR than P-Conv only networks with a few more parameters and Multi-Adds. We further improve the performance significantly with self-residual operation as presented in Table~\ref{table:ImprovedDSC}. Thus, we can conclude that our Self-Residual DSC can overcome the defects of the depthwise convolution with a simple residual with barely no additional computation costs and parameters. 


\begin{table}[t]
\begin{center}
\caption{The experiment results of different combination of pooling layers and upsamplers. The results are generated with cosine scheduler on Set5 with $\times$2 scale. }
\scalebox{0.8}{
\begin{tabular}{|l|c|c|c|c|c|c|c|}
\hline
Nearest  & \XSolid  & \Checkmark &  \Checkmark & \XSolid & \Checkmark & \XSolid \\
\hline
Bi-linear & \Checkmark & \XSolid   & \XSolid & \Checkmark & \XSolid &  \Checkmark  \\
\hline
Max-pooling & \XSolid & \XSolid & \Checkmark & \Checkmark  & \XSolid  & \XSolid\\
\hline
Avg-pooling & \Checkmark & \Checkmark & \XSolid & \XSolid & \XSolid  & \XSolid\\
\hline
PSNR & 37.83  & 37.82 & 37.85 & 37.87 & 37.82 & 37.83   \\
\hline
\end{tabular}
}
\label{table:PUDablation}
\vspace{-3mm}
\end{center}
\end{table}

\noindent\textbf{Effectiveness of the pixel-unshuffled downsampling.}
We have run six experiments to target the best combination of the pooling layer and the upsampler. The results are presented in Table~\ref{table:PUDablation}. From the table, we can observe that the model with max-pooling layer and bi-linear upsampling can achieve the best performance among all combinations.

We also compare the performance of the pixel-unshuffled downsampling with other kinds of downsampling operations. The results are shown in Figure~\ref{fig:Kernel_size}. We compare the PUD module, depthwise convolution with 2 strides, max pooling, and average pooling with both bi-linear and nearest upsampling. As shown in the figure, the pixel-unshuffled downsampling is far beyond others.

Besides, we construct a network with only pixel-unshuffled downsampling and compare its performance with the networks constructed by the baseline DSC, the Self-Residual DSC, and the PUB. The results are shown in Table~\ref{table:ImprovedDSC}
The network constructed with only pixel-unshuffled downsampling performs far worse than the baseline DSC. However, the network of PUB can performs better than the Self-Residual DSC. Specifically, the network constructed by PUB achieves $0.03$dB higher than the Self-Residual DSC only. The results show that the pixel-unshuffled downsampling can enhance the performance, but it only works together with other spatial operations.

\begin{figure}[!t]
\setlength{\abovecaptionskip}{-0.0cm}
\setlength{\belowcaptionskip}{-0.2cm}
\centering
\includegraphics[width=0.8\linewidth]{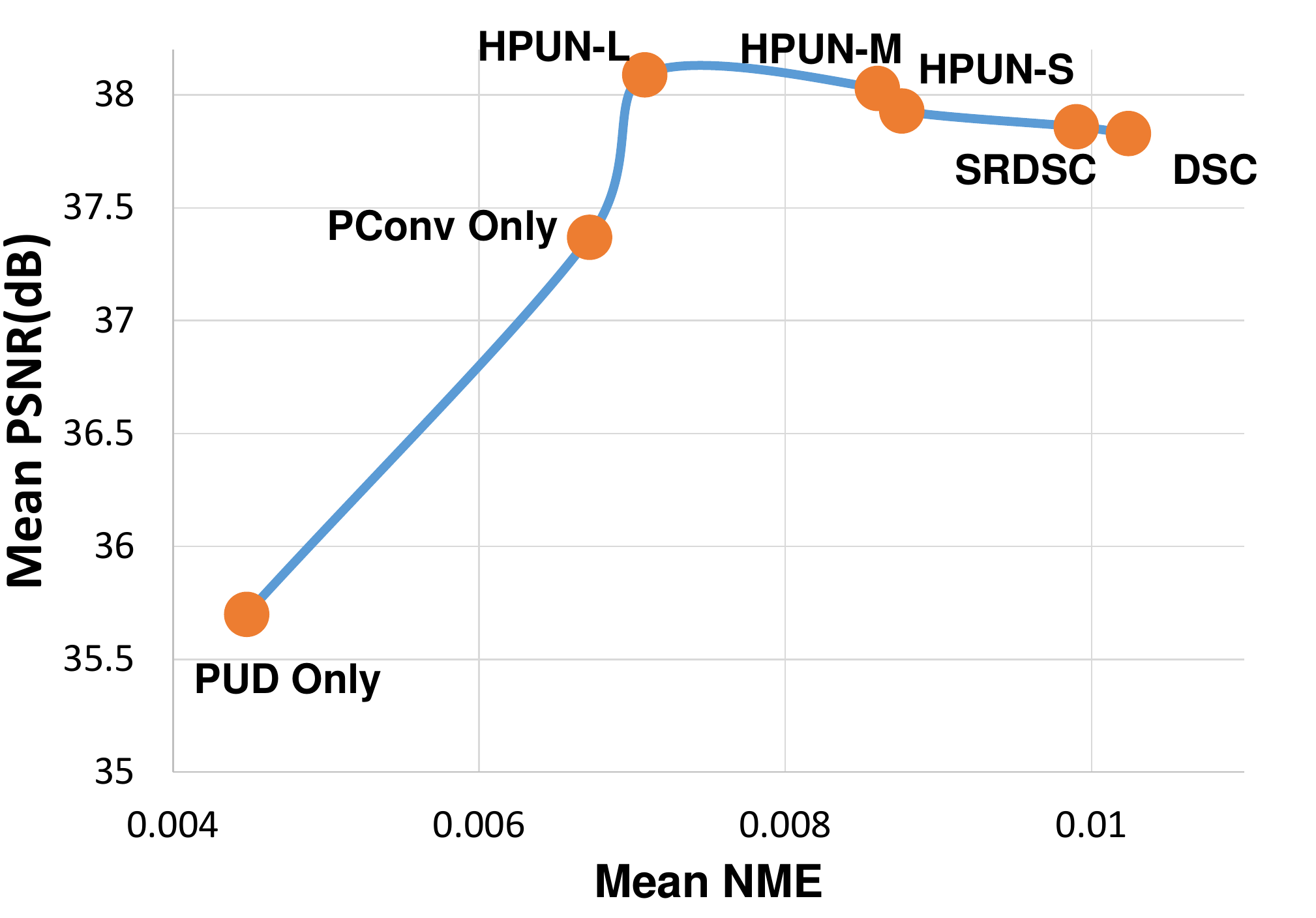}
\caption{Charts of the relationship between PSNR and the NME. The chart is generated on the whole Set5 dataset with $\times2$ scale. Abbrevs.: PUD=Pixel-Unshuffled Downsampling, SRDSC=Self-Residual DSC.
}
\label{fig:PSNR_NME_All}
\end{figure}

\begin{figure}[!t]
\setlength{\abovecaptionskip}{-0.0cm}
\setlength{\belowcaptionskip}{-0.2cm}
\centering
\includegraphics[width=0.9\linewidth]{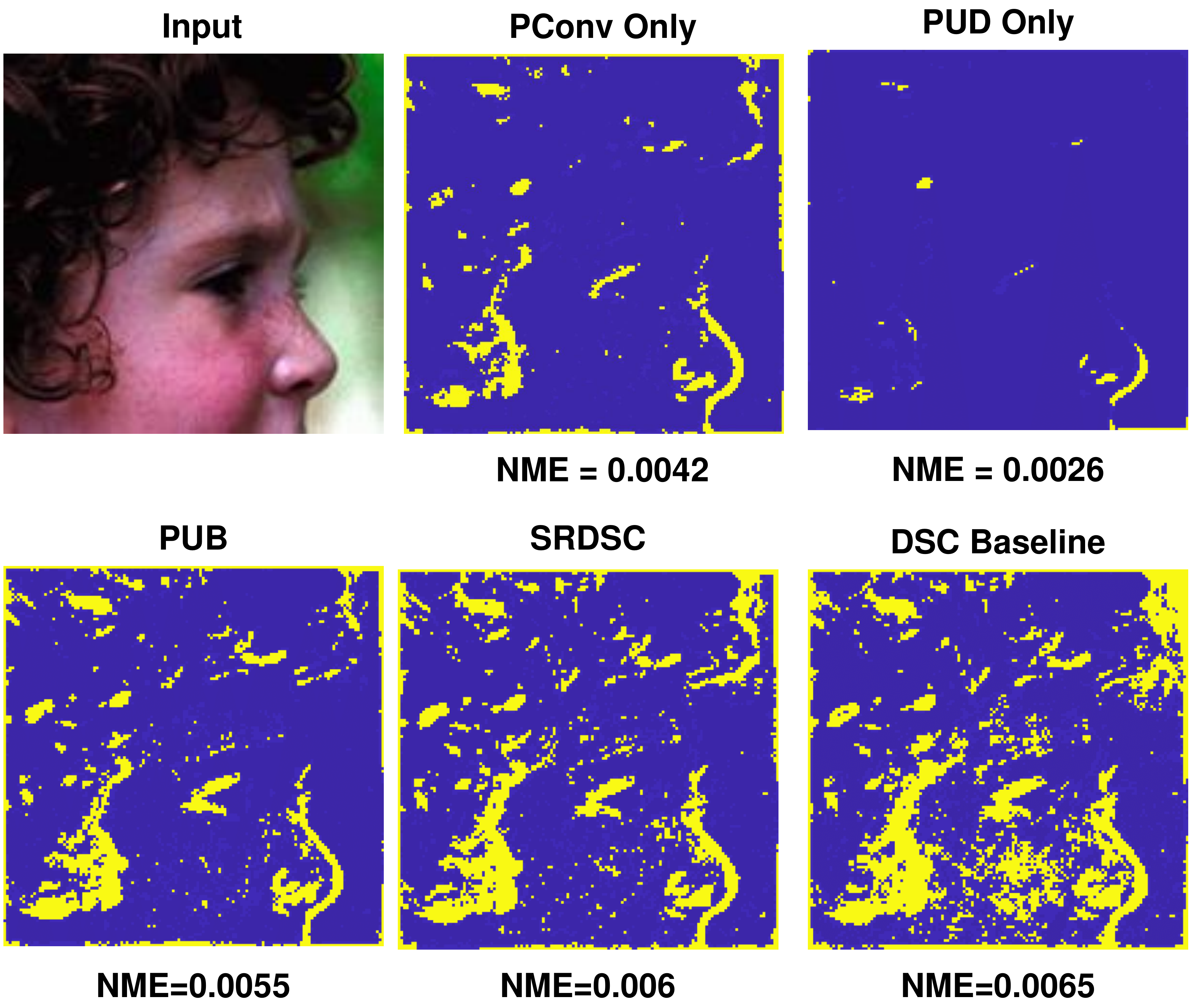}
\caption{The NME visualization of Pointwise Convolution, DSC, Self-Residual DSC, pixel-unshuffled downsampling, PUB. The results are generated on Set5 dataset with $\times2$ scale. The brighter color means the higher NME value. We binarize the features for better visualization.  Abbrevs.: PUD=Pixel-Unshuffled Downsampling, SRDSC=Self-Residual DSC. Better view in color.
}
\label{fig:Intuition}
\vspace{-5mm} 
\end{figure}

\begin{table*}[h!]
\small
\centering
\begin{center}
\caption{Benchmark results with \textbf{BI} degradation. We show average PSNR \& SSIM values for scaling factor $\times4$. We highlight the top-2 least params, Multi-Adds, and performance on each datasets with \textbf{\textcolor{red}{bold red}} (best) and \underline{\textcolor{blue}{underlined blue}} (second), respectively. + denotes the results are generated with self-ensemble.}
\scalebox{0.75}{

\begin{tabular}{|l |c|c|c|c|c|c|c|c|c|c|c|c|c|c|}
\hline
\multirow{2}{*}{Method Name} & \multirow{2}{*}{Scale} & \multirow{2}{*}{Params} & \multirow{2}{*}{Multi-Adds} & \multicolumn{2}{c}{Set5} & \multicolumn{2}{|c}{Set14} & \multicolumn{2}{|c}{B100} & \multicolumn{2}{|c}{Urban100} & \multicolumn{2}{|c|}{Manga109}\\
\cline{5-14}
 & & & & PSNR & SSIM & PSNR & SSIM & PSNR & SSIM & PSNR & SSIM & PSNR & SSIM \\
\hline
\hline
Bicubic & $\times 4$ & - & - & 28.42 & 0.8104 & 26.00 & 0.7027 & 25.96 & 0.6675 & 23.14 & 0.6577 & 24.89 & 0.7866 \\
SRCNN\cite{dong2016image} & $\times 4$ & \textbf{\textcolor{red}{57K}} & 52.7G & 30.48 & 0.8628 & 27.50 & 0.7513 & 26.90 & 0.7101 & 24.52 & 0.7221 & 27.58 & 0.8555\\

VDSR\cite{kim2016accurate} & $\times 4$ & 665K & 612.6G & 31.35 & 0.8830 & 28.02 & 0.7680 & 27.29 & 0.7260 & 25.18 & 0.7540 & 28.83 & 0.8870\\

DRCN\cite{kim2016deeply} & $\times 4$ & 1774K & 9788.7G & 31.53 & 0.8854 & 28.02 & 0.7670 & 27.23 & 0.7233 & 25.18 & 0.7524 & 28.93 & 0.8854\\

DRRN\cite{tai2017image} & $\times 4$ & 297K & 6797.0G & 31.68 & 0.8888 & 28.21 & 0.7720 & 27.38 & 0.7284 & 25.44 & 0.7638 & 29.46 & 0.8960\\

LapSRN\cite{lai2017deep} & $\times 4$ & 813K & 149.4G & 31.54 & 0.8850 & 28.19 & 0.7720 & 27.32 & 0.7280 & 25.21 & 0.7560 & 29.09 & 0.8845\\

MemNet\cite{tai2017memnet} & $\times 4$ & 677K & 623.9G & 31.74 & 0.8893 & 28.26 & 0.7723 & 27.40 & 0.7281 & 25.50 & 0.7630 & 29.42 & 0.8942\\

CARN\cite{CARN2018} & $\times 4$ & 1592K & 90.9G & 32.13 & 0.8937 & 28.60 & 0.7806 & 27.58 & 0.7349 & 26.07 & 0.7837 & 30.47 & 0.9084\\

IDN\cite{IDN2018} & $\times 4$ & 553K & 32.3G & 31.82 & 0.8903 & 28.25 & 0.7730 & 27.41 & 0.7297 & 25.41 & 0.7632 & 29.41 & 0.8942\\

SRFBN-S\cite{SRFBN-S2019} & $\times 4$ & 483K & 852.9G & 31.98 & 0.8923 & 28.45 & 0.7779 & 27.44 & 0.7313 & 25.71 & 0.7719 & 29.91 & 0.9008\\

IMDN\cite{hui2019lightweight} & $\times 4$ & 715K & 40.9G & 32.21 & 0.8948 & 28.58 & 0.7811 & 27.56 & 0.7353 & 26.04 & 0.7838 & 30.45 & 0.9075 \\


LatticeNet\cite{latticenet2020} & $\times 4$ & 777K & 43.6G & 32.18 & 0.8943 & 28.61 & 0.7812 & 27.57 & 0.7355 & 26.14 & 0.7844  & - & - \\

SMSR\cite{SMSR2021} & $\times 4$ & 1006K & 41.6G & 32.12 & 0.8932 & 28.55 & 0.7808 & 27.55 & 0.7351 & 26.11 & 0.7868 & 30.54 & 0.9085 \\
LatticeNet-CL\cite{luo2022lattice} & $\times 4$ & 777K & 43.6G & 32.30 & 0.8958 & 28.65 & 0.7822 & 27.59 & 0.7365 & 26.19 & 0.7855  & - & - \\
\hline
HPUN-S  & $\times 4$ & \underline{\textcolor{blue}{246K}} & \textbf{\textcolor{red}{12.7G}} & 32.09 & 0.8931 & 28.52 & 0.7797 & 27.54 & 0.7348 & 25.86 & 0.7788 & 30.21 & 0.9043 \\
HPUN-M  & $\times 4$ & 511K & \underline{\textcolor{blue}{27.7G}} & 32.24 & 0.8950 & 28.66 & 0.7828 & 27.60 & 0.7371 & 26.12 & 0.7878 & 30.55 & 0.9089 \\

HPUN-L  & $\times 4$ & 734K & 39.7G & 32.38 & 0.8969 & 28.72 & \underline{\textcolor{blue}{0.7847}} & \underline{\textcolor{blue}{27.66}} & \underline{\textcolor{blue}{0.7393}} & \underline{\textcolor{blue}{26.36}} & \underline{\textcolor{blue}{0.7947}} & \underline{\textcolor{blue}{30.83}} & \underline{\textcolor{blue}{0.9124}} \\
\hline
LatticeNet+\cite{latticenet2020} & $\times 4$ & 777K & 348.8G & 32.30 & 0.8962 & 28.68 & 0.7830 & 27.62 & 0.7367 & 26.25 & 0.7873  & - & - \\
LatticeNet-CL+\cite{luo2022lattice} & $\times 4$ & 777K & 348.8G & \underline{\textcolor{blue}{32.39}} & \underline{\textcolor{blue}{0.8973}} & 28.71 & 0.7837 & 27.64 & 0.7375 & 26.29 & 0.7890  & - & - \\
HPUN-S+  & $\times 4$ & \underline{\textcolor{blue}{246K}} & 101.6G & 32.19 & 0.8947 & 28.63 & 0.7817 & 27.59 & 0.7360 & 25.97 & 0.7817 & 30.46 & 0.9073 \\
HPUN-M+  & $\times 4$ & 511K & 221.6G & 32.35 & 0.8964 &\underline{\textcolor{blue}{ 28.74}} & 0.7842 & 27.65 & 0.7384 & 26.24 & 0.7904 & 30.83 & 0.9117 \\

HPUN-L+  & $\times 4$ & 734K & 317.6G & \textbf{\textcolor{red}{32.47}} & \textbf{\textcolor{red}{0.8980}} & \textbf{\textcolor{red}{28.82}} & \textbf{\textcolor{red}{0.7864}} & \textbf{\textcolor{red}{27.71}} & \textbf{\textcolor{red}{0.7404}} & \textbf{\textcolor{red}{26.48}} & \textbf{\textcolor{red}{0.7974}} & \textbf{\textcolor{red}{31.09}} & \textbf{\textcolor{red}{0.9146}} \\
\hline
\end{tabular}}

\label{tab:results_BI_5sets}
\end{center}
\vspace{-5mm}

\end{table*}



\begin{figure*}[t]
\scriptsize
\centering
\begin{tabular}{cc}
\\
\hspace{-5.7mm}
\begin{adjustbox}{valign=t}
\begin{tabular}{c}
\includegraphics[width=0.2225\textwidth]{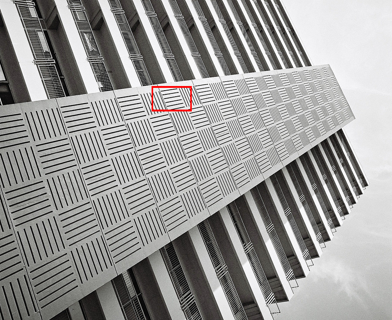}
\\
img\_092 ($\times$4)
\end{tabular}
\end{adjustbox}
\hspace{-3.3mm}
\begin{adjustbox}{valign=t}
\begin{tabular}{cccccc}
\includegraphics[width=0.150\textwidth]{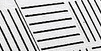} \hspace{-3.0mm} &
\includegraphics[width=0.150\textwidth]{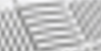} \hspace{-3.0mm} &
\includegraphics[width=0.150\textwidth]{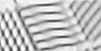} \hspace{-3.0mm} &
\includegraphics[width=0.150\textwidth]{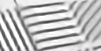} \hspace{-3.0mm} &
\includegraphics[width=0.150\textwidth]{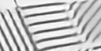} \hspace{-3.0mm}
\\
HQ \hspace{-3.0mm} &
Bicubic \hspace{-3.0mm} &
SRCNN\hspace{-3.0mm} &
VDSR\hspace{-3.0mm} &
LapSRN\hspace{-3.0mm}
\\
\includegraphics[width=0.150\textwidth]{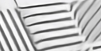} \hspace{-3.0mm} &
\includegraphics[width=0.150\textwidth]{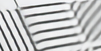} \hspace{-3.0mm} &
\includegraphics[width=0.150\textwidth]{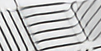} \hspace{-3.0mm} &
\includegraphics[width=0.150\textwidth]{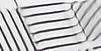} \hspace{-3.0mm} &
\includegraphics[width=0.150\textwidth]{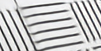} \hspace{-3.0mm}  
\\ 
MemNet\hspace{-3.0mm} &
CARN \hspace{-3.0mm} &
IMDN\hspace{-3.0mm} &
SMSR\hspace{-3.0mm} &
HPUN-L\hspace{-3.0mm}
\\
\end{tabular}
\end{adjustbox}
\\
\hspace{-6.7mm}
\begin{adjustbox}{valign=t}
\begin{tabular}{c}
\includegraphics[width=0.2225\textwidth]{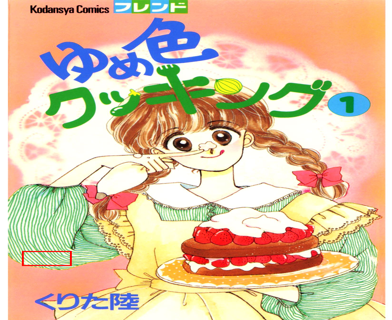}
\\
YumeiroCook. ($\times$4)
\end{tabular}
\end{adjustbox}
\hspace{-3.3mm}
\begin{adjustbox}{valign=t}
\begin{tabular}{cccccc}
\includegraphics[width=0.150\textwidth]{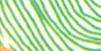} \hspace{-3.0mm} &
\includegraphics[width=0.150\textwidth]{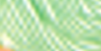} \hspace{-3.0mm} &
\includegraphics[width=0.150\textwidth]{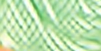} \hspace{-3.0mm} &
\includegraphics[width=0.150\textwidth]{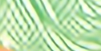} \hspace{-3.0mm} &
\includegraphics[width=0.150\textwidth]{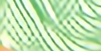} \hspace{-3.0mm}
\\
HQ \hspace{-3.0mm} &
Bicubic \hspace{-3.0mm} &
SRCNN\hspace{-3.0mm} &
VDSR\hspace{-3.0mm} &
LapSRN\hspace{-3.0mm}
\\
\includegraphics[width=0.150\textwidth]{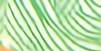} \hspace{-3.0mm} &
\includegraphics[width=0.150\textwidth]{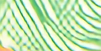} \hspace{-3.0mm} &
\includegraphics[width=0.150\textwidth]{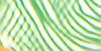} \hspace{-3.0mm} &
\includegraphics[width=0.150\textwidth]{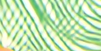} \hspace{-3.0mm} &
\includegraphics[width=0.150\textwidth]{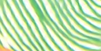} \hspace{-3.0mm}  
\\ 
MemNet\hspace{-3.0mm} &
CARN\hspace{-3.0mm} &
IMDN\hspace{-3.0mm} &
SMSR \hspace{-3.0mm} &
HPUN-L \hspace{-3.0mm}
\\
\end{tabular}
\end{adjustbox}

\end{tabular}
\vspace{-2mm}
\caption{Visual comparison with lightweight SR networks on Urban100 and Manga109.}
\label{fig:srp_visual_result_SRBIX4_lightweight}
\vspace{-3mm}
\end{figure*}

\noindent\textbf{Intuition.} For the further exploration, we generate the heatmap features using the Normalized Mean Error (NME) among their shallow features and deep features. The NME can be described as $\text{NME}=\frac{1}{N}\left \|\mathbf{{F}}^{S}-\mathbf{F}^{D}\right \|_{F}$, where $N$ means the total number of the elements in the features, $\mathbf{{F}}^{S}$ means the output features from the head block, $\mathbf{{F}}^{D}$ means the output features from the body block, and $ ||\cdot ||_{F}$ denotes the Frobenius norm. 
The NME of shallow features and deep features can be used to describe the difference between the final output and the input. Intuitively, we expect the difference to be a appropriate value: high NME means there are too many artifacts; low NME means the output is not enhanced. 
To validate the intuition, we plot the relationship between the PSNR and NME for the network constructed with pointwise convolutions, the network with DSC, the network with Self-Residual DSC, the network with pixel-unshuffled downsampling, HPUN-S, HPUN-M, and HPUN-L. The results are presented in Figure~\ref{fig:PSNR_NME_All}. We also show the result of single image in our supplementary material.

From the figure, we can find out that there exist an optimal NME to achieve the best PSNR for the similar architecture. Specifically, the NME of the pointwise structured network is much lower than the DSC baseline network, which means the main change of the features is brought by the depthwise convolution. The PSNR results of the pointwise structured network and pixel-unshuffled downsampler structured network show that limited changes will decrease the performance. However, the DSC network has large NME but limited PSNR, which means the larger NME does not represent the better performance. Although the chart generated from the bird indicate that there exist some variance, the conclusion still stands. We also visualize the NME result among the head features and body features of these networks, respectively. The visualization results are shown in Figure~\ref{fig:Intuition}. More visualization results of different models on different images are shown in our supplementary material.

Thus, we can conclude that \textbf{the PUD can significantly reduce the NME among the shallow features and deep features.} Adjusting the number of the modules will improve the performance of the architecture. Besides, we find out that the NME among the head and body features gets smaller when we integrate the standard convolution into the PUB comparing the heatmap features of HPUN-S and HPUN-M. Considering the NME of the pointwise structured network, it is natural to think about the communication among the 
features can also help to learn the similarity among the features.

From the scatter figure of mean results, we notice that the performance increases explosively with the increase of the NME at first. Then the performance starts to drop after the NME surpasses the value around $0.007$. Besides, the NME gets smaller by increasing the number of pixel-unshuffled downsamplers. Therefore, we can conclude that there may exist an optimal NME value, and increasing pixel-unshuffled downsamplers or adding residuals will reduce the NME of the network towards the optimal. The intuition can help us to design the network structure or apply the pruning strategy on SISR. However, we notice that the optimal NME is variable with the inputs from the figures in the supplementary material. We still need more experiments to validate the conclusion for different tail structures and datasets. 

\vspace{-2mm}
\subsection{Comparison Results}

To simulate LR images in SR settings, the BI degradation model is commonly used. For the BI degradation model, we compare our HPUN network to 13 cutting-edge SR methods: SRCNN~\cite{dong2016image}, VDSR~\cite{kim2016accurate}, DRCN~\cite{kim2016deeply}, DRRN~\cite{tai2017image}, LapSRN~\cite{lai2017deep}, MemNet~\cite{tai2017memnet}, CARN~\cite{CARN2018}, IDN~\cite{IDN2018}, SRFBN-S~\cite{SRFBN-S2019}, IMDN~\cite{hui2019lightweight}, LatticeNet~\cite{latticenet2020}, SMSR~\cite{SMSR2021}, and LatticeNet-CL~\cite{luo2022lattice}. All of them are popular lightweight SR methods. Note that based on the work of LatticeNet-CL~\cite{luo2022lattice}, the result in LatticeNet~\cite{latticenet2020} is generated with self-ensemble. Thus in Table~\ref{tab:results_BI_5sets} we report the results cited from LatticeNet-CL~\cite{luo2022lattice} for fair comparison.

\noindent\textbf{Quantitative Results.} Quantitative results are shown in Table~\ref{tab:results_BI_5sets} for $\times4$ SR. Among all methods, our HPUN-L achieves the new SOTA performance. Its performance on Set14, B100, and Urban100 is even better than the LatticeNet-CL+, which is generated with self-ensemble. With the self-ensemble technique, our HPUN-L achieves around $0.11$dB in average better than the LatticeNet-CL+.
Our HPUN-M can achieve top-3 performance on Set5, Set14, B100 and Manga109 with self-ensembled results excluded. Besides, compared with other competitive methods such as IMDN~\cite{hui2019lightweight}, LatticeNet~\cite{latticenet2020}, and SMSR~\cite{SMSR2021}, it only has two thirds or even fewer parameters and Multi-Adds. Furthermore, we can significantly improve the HPUN-M network using the self-ensemble technique. 
Our HPUN-S achieve comparable performance with the second least parameters among all the methods, and takes the minor computation costs. 

We specifically compare the HPUN-S network with the SRFBN-S\cite{SRFBN-S2019} and CARN~\cite{CARN2018}, since SRFBN-S also uses the low-frequency features to enhance the inference features and CARN implements group convolutions for the lightweight purpose as well. 
Our HPUN-S network can achieve comparable performance with $64.4\%$ and $4.2\%$ parameters of SRFBN-S and CARN respectively. Besides, it only takes $1.4\%$ and $13.9\%$ Multi-Adds of SRFBN-S and CARN respectively. 

\vspace{1mm}
\noindent\textbf{Visualization Results.} Visualization results are shown in Figure~\ref{fig:srp_visual_result_SRBIX4_lightweight}. The results are generated with $\times4$ scale on Urban100 and Manga109. Compared with other methods, our HPUN-L network generates better reconstruction results, especially on Manga109. As shown in the figure, the results of our HPUN-L network have fewer artifacts.

\vspace{-3mm}
\section{Conclusions}
In this paper, we proposed a lightweight network named Hybrid Pixel-Unshuffled Network (HPUN) for image SR. Specifically, we design the Self-Residual Depthwise Separable Convolution to overcome the defects of the depthwise convolution, and the pixel-unshuffled downsampling to enhance the performance with low-frequency representations. Both proposed modules take limited computation costs and parameters. With the two proposed modules, we design a lightweight block named Hybrid Pixel-Unshuffled Block with the standard convolution layer and the proposed Pixel-Unshuffled Block. The HPUN can achieve new SOTA performance with limited parameters and Multi-Adds. 
We also discover and discuss the relationship between the PSNR and the NME among the shallow features and deep features. We believe that the phenomenon should be general, and we can take advantage of it for network design.

{\small
\balance
\bibstyle{aaai23}
\bibliography{aaai23}
}

\end{document}